\documentclass[fleqn,usenatbib]{mnras}
\newcommand{\m}{10}
\newcommand{\g}{100}
\newcommand{\et}{5}
\newcommand{\h}{0.05}
\newcommand{\ep}{0.2}
\usepackage{newtxtext,newtxmath}
\usepackage{comment}

\usepackage[T1]{fontenc}
\usepackage{ae,aecompl}

\usepackage{graphicx}	
\usepackage{amsmath}	
\usepackage{amssymb}	
\usepackage{color}

\newcommand{\EQ}[1] {equation~(\ref{#1})}
\newcommand{\SEC}[1] {Section~\ref{#1}}
\newcommand{\FIG}[1] {Figure~\ref{#1}}

\def\cmpc{\mathrm{cm}^{-3}\, \mathrm{pc}}
\def\radm{\mathrm{rad}\, \mathrm{m}^{-2}}

\def\ergs{\mathrm{erg}\, \mathrm{s}^{-1}}

\title[Clumpy jet model of FRB]{Clumpy jets from black hole-massive star binaries as engines of Fast Radio Bursts}

\author[Yi Shu-Xu et al.]{
Shu-Xu Yi,$^{1,2}$\thanks{E-mail: shuxuyi@gmail.com}
K. S. Cheng,$^{1}$
Rui Luo,$^{3,4}$ \\
$^{1}$ Pokfulam Road, Department of Physics, the University of Hong Kong, Hong Kong\\
$^{2}$ Department of Astrophysics, Radboud University Nijmegen, P.O. Box 9010, NL-6500 GL Nijmegen, The Netherlands\\
$^{3}$ Department of Astronomy, School of Physics, Peking University, Beijing 100871, China \\
$^{4}$ Kavli Institute for Astronomy and Astrophysics, Peking University, Beijing 100871, China 
} 

\date{Accepted XXX. Received YYY; in original form ZZZ}

\pubyear{2018}

\begin{document}
\label{firstpage}
\pagerange{\pageref{firstpage}--\pageref{lastpage}}
\maketitle

\begin{abstract}
We propose a new model of Fast Radio Bursts (FRBs) based on stellar mass black hole-massive star binaries. We argue that the inhomogeneity of the circumstellar materials or/and the time varying wind activities of the stellar companion will cause the black hole to accrete at a transient super-Eddington rate. The collision among the clumpy ejecta in the resulted jet could trigger plasma instability. As a result, the plasma in the jet will emit coherent curvature radiation. When the jet cone aims toward the observer, the apparent luminosity can be $10^{41}-10^{42}\,\ergs$. The duration of the resulted flare is $\sim$ millisecond. The high event rate of the observed non-repeating FRBs can be explained. A similar scenario in the vicinity of a supermassive black hole can be used to explain the interval distribution of the repeating source FRB121102 qualitatively.
\end{abstract}

\begin{keywords}
radiation mechanisms: general, stars: black holes, stars: binaries
\end{keywords}



\section{Introduction}

\begin{figure}
\centering
\includegraphics[width=8cm]{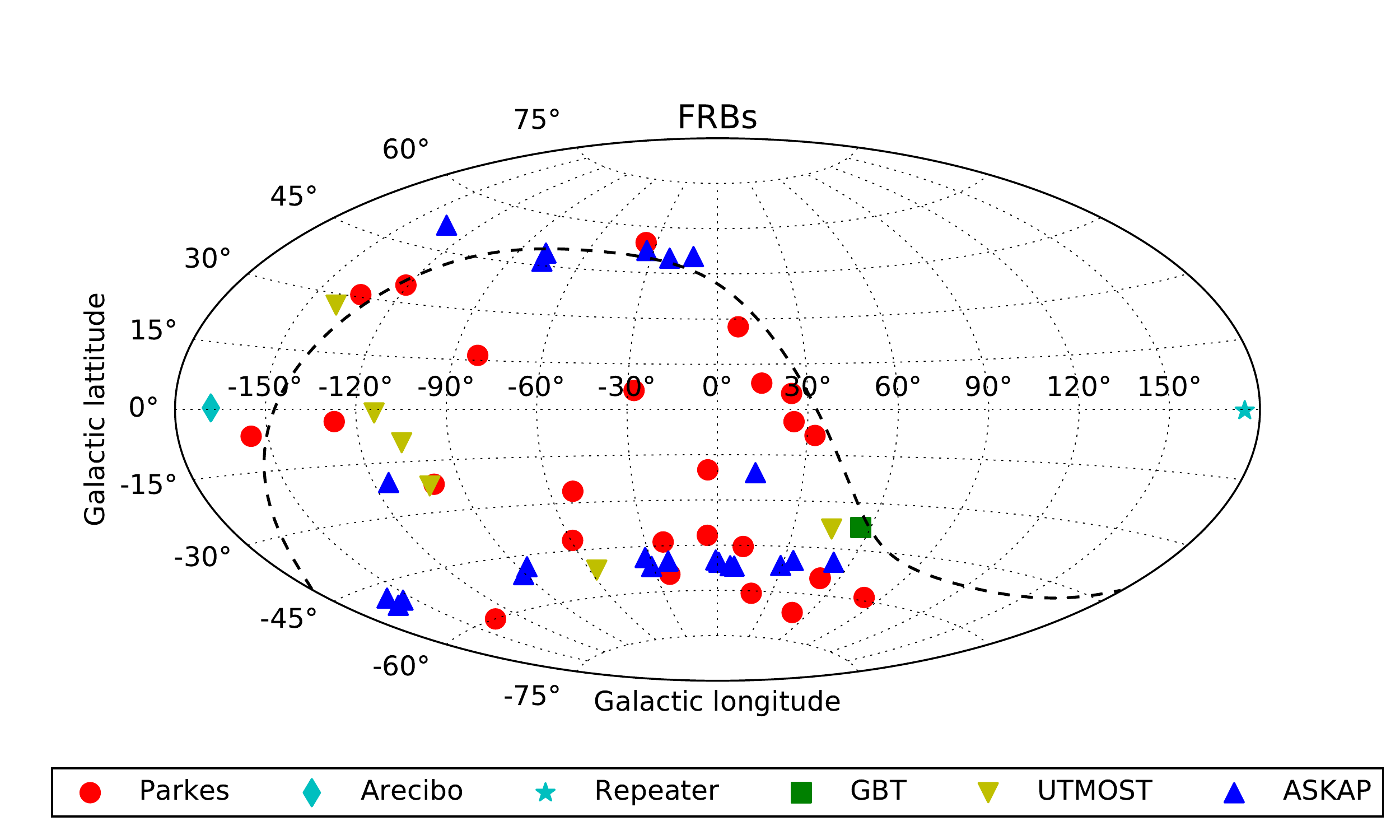}
\caption{An Aitoff projected map of current FRBs in the Galactic coordinates, in which the black dashed line denotes the celestial equator. The detections are from these telescopes: Parkes radio telescope (red circles), Arecibo radio telescope (cyan thin diamand), Green Bank Telescope (green square), UTMOST (yellow down-pointing triangles), ASKAP (blue up-pointing triangles). In particular, the repeater (FRB~121102) is marked by cyan star.}
\label{fig:skymap}
\end{figure}

High time-resolution universe contains a lot of unknowns to be revealed. As a new class of astronomical transients, Fast Radio Bursts (FRBs), exhibit basic observational features with short duration ($\sim$ 0.1 -- 10 ms), large flux density (0.1 -- 100~Jy) and prominent dispersion. From its first discovery \citep{2007Sci...318..777L} to present, over 50 such flashes have been reported as detections by Oct.\,2018, which are nearly isotropically distributed in the full sky (see \FIG{fig:skymap}, of which the data comes from \texttt{FRBCAT}\footnote{http://frbcat.org} \citep{Petroff16PASA}). FRBs are believed to have extragalactic or cosmological origin because of their large dispersion measure (DM, from about 170 to 2600 $\cmpc$). Due to remarkable flux observed, the intrinsic luminosities are estimated to be very high with a broad range ($10^{41}-10^{44}\,\ergs$, \citealt{Luo2018MN}). 

Although there are a substantial number of FRBs reported, the nature of FRBs is still a mystery. So far, theoretical models with number much larger than FRBs themselves have been proposed, related to a wide range of astronomical objects, such as pulsars \citep{Connor16MNRAS, Dai16ApJ}, magnetars \citep{PP13arXiv, Pen15ApJ, Metzger17ApJ, Belo17ApJ}, neutron stars \citep{Totani13PASJ, FR14A&A, Zhang14ApJ, Cordes16MNRAS, Zhang17ApJ}, quark stars \citep{Shand16RAA, Wang18ApJ}, white dwarfs \citep{Kashiyama13ApJ, Gu16ApJ}, black holes (BH, \citealt{Romero16PhRvD, Liu16ApJ, Katz17MNRAS}), etc..

The only repeating FRB, FRB~121102 \citep{Spitler16Natur}, draws many attentions owing to continuous observational progresses. Its host galaxy was identified as a dwarf galaxy at a redshift of 0.193 with the Very Large Array (VLA) \citep{Chat17Natur, Tend17ApJ} and a persistent radio source is found to be spatially associated with these bursts \citep{Marcote17ApJ}. At one time, people tended to think the persistent radio source could be pulsar wind nebula \citep{Belo17ApJ, Dai17ApJ} and emits via synchrotron radiation \citep{Waxman17ApJ}. However, new observations on the repeater revealed its huge Faraday rotation measure ($>10^5\,\radm$), which suggested that the highly magnetized environment is compatible with those of an accreting massive BH in the galaxy centre \citep{Mich18Natur}. The FRB progenitor is proposed to be a neutron star orbiting the central black hole \citep{Zhang18ApJ} or a young millisecond-magnetar in the supernova remnant \citep{Metzger17ApJ}. 

In this paper, we focus on another possible mechanism related to a much larger population: stellar-mass BH-massive star binaries. When the mass transfer rate from the stellar companion suddenly increases, a transient super-Eddington accretion will occur, analogous to the process in the tidal disruption event (TDE; \citealt{2018ApJ...859L..20D}) . A clumpy jet could be thus launched. We find that collision among the ejecta and coherent curvature radiation of the plasma in the jet can produce radio flares which coincides with the observational characteristics of FRBs. 

This paper is organized as follows. We present our model in \SEC{sec:mod} and we give discussion in \SEC{sec:dis}. 

\section{The model}\label{sec:mod}
We focus our interests on the situation where the materials from a massive star are being accreted onto the BH before a steady state accretion disc established. Such a situation is analogous to the transient accretion in TDEs. The materials are accreted at a nearly free-falling rate which is super-Eddington \citep{2015MNRAS.454L...6M}. The super-Eddington accretion could trigger a jet via Blandford-Znajek mechanism \citep{1977MNRAS.179..433B}. Since the accretion flow is not in a steady state, the ejecta in the jet are expected to be inhomogeneous, or in other words, clumpy \citep{2012MNRAS.423.3083M,2018ApJ...859L..20D}. When two clumps of ejecta collide with other each, plasma oscillation is activated and charge-separated bunches are formed. These bunches slide along the spiral magnetic field lines in the jet and radiate via curvature radiation. When the frequency of the curvature radiation matches the plasma frequency, the two-stream instability will grow and the charge bunch will emit coherently in a short time. \FIG{fig:model} is an illustration for the model. 
\subsection{Collision between clumpy ejecta in the jet}
We start with study of the collision between the successive clumps in the jet. 
\begin{figure}
\centering
\includegraphics[width=8cm]{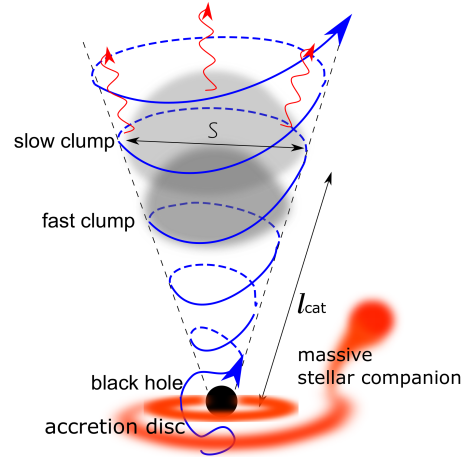}
\caption{\textbf{The illustration of the scenario.} The blue curves with arrow represent the spiral magnetic field lines in the jet. Two clumps in gray are ejecta sliding along the magnetic field lines and emitting via curvature radiation. $l_{\rm{cat}}$ is the distance from the BH to the location where the two clumps collide, where the plasma oscillation is thus triggered.}
\label{fig:model}
\end{figure} 
The Lorentz factor of the clumpy ejecta in the jet can be estimated in the following way. The energy of an ejecta with the mass $\delta M$ is: 
\begin{equation}
\gamma\delta Mc^2=\epsilon\dot{M}c^2\delta t,
\label{eqn:gammadeltaM}
\end{equation}
where $\delta t$ is the interval between the successive ejecta, $\dot{M}$ is the accretion rate onto the black hole and $\epsilon$ is the energy converting efficiency from the accretion to the jet. $\dot{M}$ is scaled with the Eddington accretion rate:
\begin{equation}
\dot{M}=\eta\dot{M}_{\rm{Edd}}.
\label{eqn:edding}
\end{equation}
The typical dimension of the clumps corresponds to the height of the accretion flow at the inner edge, $h$. Therefore the mass of the clump can be approximated as:
\begin{equation}
\delta M=h^3\rho_0,
\label{eqn:deltaM}
\end{equation}
where $\rho_0$ is the density of the material at the inner edge of the accretion flow. The accretion rate is related to the properties of the accretion flow as:
\begin{equation}
\dot{M}=2\pi\rho_0vr_{\rm{in}}h,
\label{eqn:Mdot}
\end{equation}
where $v$ is the inflow velocity of the materials at the inner edge $r_{\rm{in}}$. $v$ can be approximated with the free-falling velocity:
\begin{equation}
v=\sqrt{\frac{2GM_\bullet}{r_{\rm{in}}}},
\end{equation}
where $M_\bullet$ is the mass of the BH. We use the radius of the innermost stable circular orbit of the BH, $r_{\rm{in}}=3r_{\rm{s}}$, where $r_{\rm{s}}$ is the Schwarzschild radius of the BH. As a result, $v=c/\sqrt{3}$ where $c$ is the speed of light. 

For the estimation of $\gamma$, we do not need the explicit formula of $v$. We use the size of the clump $h$ as the average separation, thus the interval between ejecta is scaled with:
\begin{equation}
\delta t\sim h/v.          
\label{eqn:deltat}
\end{equation}
Taking equations (\ref{eqn:deltaM},\ref{eqn:Mdot},\ref{eqn:deltat}) into equation (\ref{eqn:gammadeltaM}), we obtain that
\begin{equation}
\gamma\sim2\pi\epsilon\frac{r_{\rm{in}}}{h},
\label{eqn:gamma}
\end{equation}
which is independent with the mass and the accretion rate of the BH. We denote the dimensionless $h/r_{\rm{in}}$ as $\tilde{h}$ in the following texts for simplicity. From equation (\ref{eqn:gamma}) we expect the $\gamma$ to take value from $\sim10$ to $\sim100$. 
Since equation (\ref{eqn:gamma}) is an approximated equation, we only take equation (\ref{eqn:gamma}) as a rough hint of the range of possible $\gamma$ and still treat $\gamma$ as an independent parameter in the following of the model.   

Since the accretion flow is inhomogeneous, the velocities among the ejecta are expected to be nonuniform. In a case where a slower clump is ejected earlier than a faster one, the latter clump will collide with the earlier clump at a catch-up distance $l_{\rm{cat}}$\footnote{when the Lorentz factors of the two clumps are of the same order of magnitudes, and the different in the Lorentz factor $\delta\gamma$ is also in the same order of magnitude}:
\begin{equation}
l_{\rm{cat}}\approx c\gamma^2\delta t.
\label{eqn:lcat}
\end{equation}
The ejecta slide along the spiral magnetic field lines, and emit electromagnetic waves via curvature radiation. The frequency of curvature radiation in the rest frame $\nu_{\rm{cur}}$ is determined by the local curvature radius of the magnetic field lines and the Lorentz factor of the plasma. Throughout the paper, we assume for simplicity that the plasma remains cold, i.e., the microscopic motion is negligible and we only consider the bulk motion. Since the magnetic field lines are highly spiral in the jet, the curvature radius of the magnetic field lines can be approximated as the radius of the jet cone (see the illustration in \FIG{fig:model}). The radius of the jet cone is denoted as $s(l)$ at $l$, therefore 
\begin{equation}
\nu_{\rm{cur}}(l)=2\gamma_\perp\gamma_\parallel^3c/s(l),
\end{equation}
where $\gamma_\parallel$ is the Lorentz factor corresponding to the sliding of the plasma along the magnetic field lines, $\gamma_\perp$ corresponds to the bulk velocity along the line of sight (in a case where the jet cone is towards the observer). For simplicity we use $\gamma_\perp\sim\gamma_\parallel\sim\gamma$. 
The frequency of the curvature radiation as function of $l$ is calculated as
\begin{equation}
\nu_{\rm{cur}}(l)=3.8\tilde{h}^{-1}\frac{\gamma_{100}^2}{m}\frac{l_{\rm{cat}}}{l}\,\text{GHz},
\label{eqn:nucur}
\end{equation}
where $m$ is the mass of the black hole in units of $M_\odot$, $\gamma_
{100}\equiv\gamma/100$. When calculating \EQ{eqn:nucur}, we use \EQ{eqn:lcat} and with the assumption that the opening angle of the jet cone is 0.1, i.e.,
\begin{equation}
s(l)=0.1l.
\end{equation}

\subsection{Coherent curvature radiation}
The collision between the successive clumps triggers the plasma oscillation, and the charges in the plasma are displaced with the plasma frequency $\nu_{\rm{pls}}$. The charge-separated plasma acts like a bunch of charged-particles, which moves across the spiraled magnetic field lines and radiates via curvature radiation. When the condition is satisfied such that $\nu_{\rm{pls}}=\nu_{\rm{cur}}$, the charged bunch will radiate coherently.

The plasma frequency in the observer's frame is:
\begin{equation}
\nu_{\rm{pls}}(l)=2\sqrt{\frac{\gamma e^2n_{\rm{e}}(l)}{\pi m_{\rm{e}}}},
\label{eqn:fpls}
\end{equation}
where $e$ and $m_{\rm{e}}$ are the charge and the mass of the electron, $n_{\rm{e}}$ is the number density of the electrons, which is
\begin{equation}
n_{\rm{e}}(l)=2\rho(l)/m_{\rm{p}},
\label{eqn:ne}
\end{equation}
where $m_{\rm{p}}$ is the mass of proton. 

\begin{figure}
\centering
\includegraphics[width=7cm]{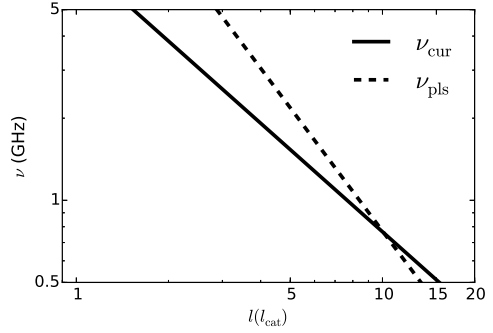}
\caption{\textbf{$\nu_{\rm{cur}}$ and $\nu_{\rm{pls}}$ as functions of $l$.} Here we use $m=\m$, $\gamma=\g$, $\eta=\et$, $\epsilon=\ep$ and $\tilde{h}=\h$. }
\label{fig:freq}
\end{figure}

Since the dimension of the clump expands along the jet with increasing radius of the jet cone $s(l)$, the density of the plasma decreases accordingly as
\begin{equation}
\rho(l)=\rho_0\left(\frac{h}{s(l)}\right)^3,
\label{eqn:rhol}
\end{equation}
and from equations (\ref{eqn:edding},\ref{eqn:Mdot}):
\begin{equation}
\rho_0=\frac{\eta\dot{M}_{\rm{Edd}}}{2\pi vr_{\rm{in}}h}.
\label{eqn:rho}
\end{equation}
The Eddington rate is explicitly expressed as:
\begin{equation}
\dot{M}_{\rm{Edd}}=\frac{1.26\times10^{38}m\,\ergs}{\epsilon c^2}.
\end{equation}
From the above equations, we obtain that
\begin{equation}
\nu_{\rm{pls}}(l)=3.45\sqrt{\frac{\eta}{m\epsilon\tilde{h}}}\gamma^{-2.5}_{100}\left(\frac{l_{\rm{cat}}}{l}\right)^{1.5}\,\text{GHz}.
\end{equation}
Note that $\nu_{\rm{pls}}$ decreases faster than $\nu_{\rm{cur}}$ as $l$ increases (See \FIG{fig:freq}). Therefore as long as $\nu_{\rm{pls}}>\nu_{\rm{cur}}$ at $l=l_{\rm{cat}}$, the condition that $\nu_{\rm{pls}}=\nu_{\rm{cur}}$ can always meet at some distance $l>l_{\rm{cat}}$. In \FIG{fig:freq} we plot an example when $m=\m$, $\gamma=\g$, $\eta=\et$, $\epsilon=\ep$ and $\tilde{h}=\h$. The corresponding frequency of the coherent radiation is $\sim1$\,GHz.  

Now we will estimate the life time of the curvature radiation. 
Let $Ze$ be the net charge in the bunched plasma, the size of which is $L$. When $\nu_{\rm{p}}\sim\nu_{\rm{cur}}=1\,$GHz and $\gamma=\g$, 
\begin{equation}
L\sim c/\nu^\prime_{\rm{p}}\sim 2c\gamma/\nu_{\rm{p}}\sim10^3\,\rm{cm}.
\end{equation}
The estimation of $Z$ follows the practice of \cite{1975ApJ...196...51R}:
\begin{equation}
ZE_{\rm{p}}=(Ze)^2/L,
\end{equation}
where $E_{\rm{p}}$ is the rest energy for each plasma particle, $m_{\rm{p}}c^2=1.5\times10^{-3}\,$erg. 
$Z$ is estimated to be $\sim10^{19}$. 
The life time of the curvature radiation is:
\begin{equation}
\tau=\frac{Z\gamma m_{\rm{p}}c^2}{\frac{2}{3}(Z^2e^2/c^3)\gamma^4(2c^2/s)^2}\sim10\gamma^{-3}_{100}\,\mu\text{s}, 
\end{equation}
which corresponds to the smallest temporal structure of the FRB. It is essential that $\tau$ is larger than the time scale of plasma instability growth, $t_{\rm{g}}=1/\nu_{\rm{p}}\sim$ns. Otherwise the plasma cannot be significantly bunched. We note from equations (14, 15), that $\tau$ is inverse proportional to $t_{\rm{g}}$. As a result, the frequency of coherent radiation should be limited to be larger than $\sim10$\,MHz. We see from equation (17) that the emission can be up to a few GHz. 

\begin{figure}
\centering
\includegraphics[width=8cm]{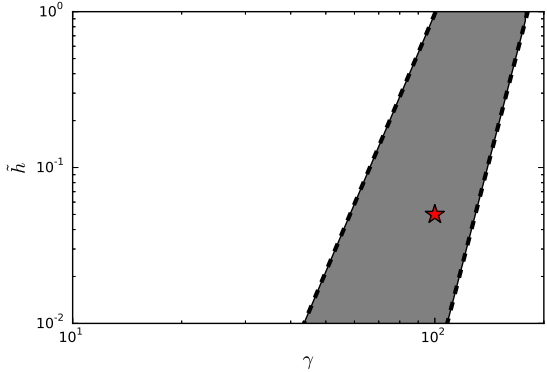}
\caption{\textbf{The parameter space for $\tilde{h}$ and $\gamma$}. The shaded region is where the coherent radio emission condition in equation (\ref{eqn:condition}) is satisfied when $m=\m$, $\epsilon=\ep$ and $\eta=\et$. The red star marks the parameters using in figure 2.}
\label{fig:params}
\end{figure}

The above argument sets a necessary condition for the coherent radio emission:
\begin{equation}
\nu_{\rm{cur}}(l)=\nu_{\rm{pls}}(l)>10\,\text{MHz,}\qquad\text{at }l>l_{\rm{cat}}.
\label{eqn:condition}
\end{equation} 
In \FIG{fig:params}, we illustrate the parameter-space of $\tilde{h}$ and $\gamma$ corresponding to equation (\ref{eqn:condition}), given $m=\m$, $\epsilon=\ep$ and $\eta=\et$. 
\subsection{Duration and the luminosity of the flare}
The time scale of the flare is determined by the duration of the super-Eddington accretion of the transient accretion disc, which corresponds to the free falling time scale near the black hole: 
\begin{equation}
\tau_{\rm{dur}}\sim r_{\rm{in}}/c.
\end{equation}
For a BH with $10\,M_\odot$, $\tau_{\rm{dur}}\approx0.3\,$ms. This time scale is in accordance with the millisecond duration of FRBs. 

The apparent luminosity is:
\begin{equation}
L=\frac{2\pi}{\Delta\Omega}\epsilon\dot{M}c^2,
\end{equation}
where $\Delta\Omega$ is the solid angle of the jet. As we assumed above, the opening angle of the jet is $\sim0.1$, therefore $\Delta\Omega\sim0.01$. The luminosity is therefore 
\begin{equation}
L=8\epsilon m\eta\times10^{40}\,\ergs.
\label{eqn:L}
\end{equation}
With typical masses of stellar BHs and super-Eddington accretion rate, $L$ can be $10^{41}-10^{42}\,\ergs$. As shown in equation (20), the efficiency of the coherent radio emission is very high. Therefore, the radio luminosity is approximately equal to total luminosity.
\subsection{Event rate}
With a flux density limit at 1\,Jy, the radio flare can be detected up to a luminosity distance of $D_{\rm{>1\,Jy}}$, where
\begin{equation}
\frac{L}{4\pi D^2_{\rm{>1\,Jy}}}\approx1\,\text{GHz}\times1\,\text{Jy}.
\end{equation}
Considering equation (\ref{eqn:L}), the luminosity distance is 
\begin{equation}
D_{\rm{1\,Jy}}\approx0.3\sqrt{\frac{2\epsilon m\eta}{\pi}}\,\text{Gpc}.
\end{equation}
An estimation of the detecting rate of the radio emission from the above-mentioned scenario is:
\begin{equation}
\mathcal{R}=\frac{n_{\rm{G}}N_{\rm{BH}}}{T}\frac{2\Delta\Omega}{4\pi}V,
\label{eqn:mathR}
\end{equation}
where $N_{\rm{BH}}$ is the averaged number of BH-massive star binaries per galaxy, $n_{\rm{G}}$ is the number density of galaxies, $T$ is the averaged recurrence time scale of the radio flares, and $V$ is the comoving volume within which the radio emission flux density are larger than the detectable limit, which is approximately $\sim\text{Gpc}^3$. The factor two in equation (\ref{eqn:mathR}) accounts for the double sides of the jet cone. 

We use $n_{\rm{G}}=0.1\,\text{Mpc}^{-3}$ as an approximation in the low red-shift universe \citep{2016ApJ...830...83C}; \cite{2016A&A...587A..61C} estimated a total population of $\sim1300$ BH transients in the Milky way. As discussed above, we expect to see the violent variation of mass transfer in wind accreting BH binaries with high mass companion. Therefore we use $N_{\rm{BH}}=100$, i.e., $10\%$ of the total estimated population for the evaluation of the event rate.

The recurrence of the radio flare depends on a fast increasing of mass transfer from the donor companion. This circumstance can occur when the BH is crossing the dense stellar disc of the companion, as in the cases of $\gamma$-ray binaries with Be stars \citep{2013A&ARv..21...64D}. In this case, the accretion rate from the circumstellar materials will increase rapidly when the BH is inside the stellar disc. The recurrence time scales of the radio flares are thus corresponding to the orbital period of the binaries, which are typically a few years; Besides, many wind-accretion binaries have been observed to have super-orbital variability\footnote{the variability with time scale longer than the orbital period} \citep{2013ApJ...778...45C}. The super-orbital periodicity are believed to related with the modulation on the mass-loss rate from the stellar companion \citep{2008MNRAS.389..608F,2006A&A...458..513K}. A boost of the mass-loss rate from the donor can also cause the increasing of the accretion rate onto the BH. Note that the parameters $(\tilde{h},\gamma,\epsilon, \eta)$ must fall in certain region given $m$, in order to satisfy the necessary condition of the coherent radio emission as discussed in section II. Therefore, the radio flare might not appear in every orbital or super-orbital period, but will recur after several periods. We use $T=10\,$yrs for an rough estimation of the recurrence time scale. 

Taking $V$, $N_{\rm{BH}}$, $n_{\rm{G}}$ and $T$ into \EQ{eqn:mathR}, the event rate are evaluated to be $\sim10^{4}\,\rm sky^{-1}\, day^{-1}$, which is consistent with the statistic studies of FRBs. Recently, FRBs are observed in a new low frequency domain $400-800\,$MHz with CHIME/FRB \citep{2018BoyleAtel}. Together with the other observations, e.g., 4-8 GHz observation of FRB121102 \citep{2018ApJ...863....2G} and \cite{2018Natur.562..386S}, these new observations indicates that FRB occurs in spectral islands that move around in frequency, therefore the event rate may be higher than the value estimated previously.
\section{Discussion}
\label{sec:dis}
\subsection{FRB 121102}
\cite{2018MNRAS.475.5109O} found the intervals between the successive bursts of FRB 121102 present the tendency of clustering around the timescale of $10^3$\,s. Although the conclusion needs further confirmation with larger data samples, this work may provide some insight of the possible underlying physical mechanism. More recently \cite{2018ApJ...863....2G} observed 23 bursts from this source in a continuous observation of 6 hours, 21 bursts of which occurred in the first 60 minutes, and 18 bursts occurred in the first 30 minutes. Therefore the highest bursts rate of this source is $1/100\,\text{s}^{-1}$. We think that the mechanism of this source is different with above-mentioned stellar mass BH-massive star model since the latter cannot produce such a high bursts rate. Existing models include the millisecond-magnetar interpretation \citep{Metzger17ApJ,2018MNRAS.481.2407M}. With the hint that FRB121102 was localized to the centre of its host galaxy, we propose that the repeating bursts are from the stellar mass BH near the supermassive BH (SMBH) in the galactic centre. The stellar mass BH orbits across the clumpy inner region of the accretion disc around the SMBH. The inhomogenousity could arise from the instability in the accretion disc. Each time when the stellar BH crosses a gas clump, a transient accreting process occurs. Unlike the stellar mass BH-massive star binaries, the accreted materials have no angular momentum. If the stellar BH is a Kerr BH, a Blandford-Znajek jet can also be launched even in this case. As a result, the coherent radio emission can be triggered similarly. After all, the luminosity of the persistent source that coincides with FRB 121102 is consistent with a low luminous accreting SMBH \citep{Chat17Natur}, and the rotation measure of the bursts also agrees with this scenario \citep{Mich18Natur}. 

Such interpretation of the repeating source implies several observational consequences. The short intervals of $\sim100-1000$\,s of the bursts corresponds to the separation between clumps. If the orbit of the stellar BH around the SMBH is highly elliptical, we expect two quiescent periods of the FRB in each orbit. When the BH is outside the accretion disc of the SMBH and around the periastron, the source undergoes a shorter quiescent period; the orbital phases outside the accretion disc and around the apastron corresponds to a longer quiescent period. Another major prediction on the repeating source is that, the outbursts of the sources could be found to clustered periodically, corresponding to the orbital period of the BH, after enough data accumulated. 
\subsection{The compact companion}
It is possible for a neutron star to accrete from its massive companion at a super-Eddington rate, as were found in the Ultra-luminous X-ray pulsars \citep{2014Natur.514..202B,2016ApJ...831L..14F,2017Sci...355..817I,2017MNRAS.466L..48I}. However, since a neutron star is high magnetized, the accretion flow is truncated at the Alfv\'en radius where the magnetic pressure of the neutron star balances the ram pressure of the infalling materials. The typical Alfv\'en radius is $\sim10^8$\,cm, which is many orders of magnitudes larger than the radius of a neutron star. As a result, the conditions of millisecond duration of the model cannot be satisfied with a neutron star companion. 
\subsection{Distribution in the host galaxies}
In this paper, we propose that FRBs originate from stellar mass BH-massive star binaries. Since massive stars are short lived, we expect such system are more populated among active star forming regions. As a result, our model predicts that FRBs should statistically concentrate towards star forming galaxies. Furthermore, the sources should distribute close to the plane of disc galaxies.   
\section*{Acknowledgement}
The authors appreciate the helpful discussion about TDE accretion with Dr. Dai, L. X.. KSC and SXY are supported by a GRF grant under 17310916.



\bsp	
\label{lastpage}
\end{document}